\documentstyle{article}

\begin{document}

\input epsf.tex

\title{
\bf Is there really no quantum form of Einstein Gravity?}
\author{ A.Y. Shiekh \\
International Centre for Theoretical
Physics, Miramare, Trieste, Italy }

\date{}

\maketitle

\begin{abstract}
\begin{center} It is well known that Einstein gravity is
non-renormalizable; \\ however this does not preclude the
existence of a quantum form.
\end{center}
\end{abstract}

\begin{quote} {\em ``Everything should be made as simple as
possible, but not simpler.''} -- Albert Einstein
\end{quote}

\baselineskip = 16pt

\section{Introduction}

\subsection{Is Einstein gravity incomplete?}

  Reviewing a subject in less than an hour is hard enough, but
reviewing one that does not exist is even harder still; for
Einstein gravity has not been quantized [reviewed in Isham,
1981]. We will be taking a rather orthodox approach in reviewing
the obstructions to the quantization of traditional Einstein
gravity. But there is virtue in sometimes looking back over lost
campaigns in life.

Despite the fact that it seems impossible to quantize, this does
not exclude the existence of a quantum form of Einstein gravity
[Shiekh, 94; 96]; in the same way as prehistoric man had a source
of fire (lightning strikes on trees), though not necessarily a
means to produce it. It is sometimes all too easy to
forget that the world is quantum, and that the classical picture
is but a special case, so we are actually working backwards when
we try to derive the more general case from a restrictive form,
and there is no reason to believe that this is always possible.
However, we are compelled to follow this route out of a lack of
choice.

\subsection{Need gravity be quantize at all?}

Is it not possible that although the other forces of nature are
seen to be quantized, that perhaps gravity, which is presently
not seen as a force at all, need not be quantized [Feynman, 1963;
Kibble, 1981]? The concept of force is a classical one that does
not even make an appearance in quantum theory, where distribution
is governed by the overlap of wave functions. Newtonian gravity
certainly carries the notion of force, while the Einstein view is
of matter in free fall in a curved space-time. This force
free picture of classical gravity might lead one to propose
leaving the curved space time unquantized, and to have the
quantum fields play out on this arena. It is an appealing
scenario, and would present a resolution to the problem of
quantization, claiming it is not necessary at all.

However the gravitational field, if left classical, could be used
to make measurements on a quantum field, and being classical
would not be subject to the Heisenberg uncertainty principle.
Thus, if one were to look at a quantum field using gravity as a
probe, one would be able to extract information about the quantum
field that defied the Heisenberg principle. Such arguments are
far from water-tight, but do make a strong case for the
quantization of gravity.

Despite the very geometric picture that is usually assigned to
Einstein gravity, it can be made to look like the other
perturbative field theories, with the graviton a spin two
particle. It is this very traditional, particle physics
perspective, that we will be following. Unfortunately, the whole
structure now takes on a very mathematical structure, and at
times it is hard to keep contact with the physics.

Actually, we will be adding a second theme to the quantization
that is not strictly necessary, namely renormalization by
analytic continuation. In these methods there is no subtraction
of infinities, but rather a reinterpretation of the formulae.
This is a formulation devoid of infinities, so completely
by-passing the problem of interpreting the divergences. At the
risk of mixing concepts, the quantization is illustrated using
this novel method, which can be shown to be equivalent, in
effect, to the older techniques.

\subsection{How far can we go without renormalizing?}
  It might come as somewhat of a surprise, but one can actually
go part the way to quantizing without having to deal with
renormalization. It was discovered [Donoghue, 1994; 1994]
that the one loop correction to the Newtonian potential does not
encounter any infinities, and so makes no demands on
renormalization.
  The best way to see this, is to use dimensional analysis to
predict the manner of corrections anticipated [Donoghue, 1995].
In configuration space this takes the form:

\begin{equation} V\left( r \right)=-{{Gm_1m_2} \over r}\left(
{1+\alpha{{G(m_1+m_2)}
\over {rc^2}}+\beta{{G\hbar } \over {r^2c^3}}+\ldots } \right)
\end{equation}

\noindent which includes the first relativistic correction, and
the first quantum correction (also relativistic). These power law
corrections come from momentum space (where we normally
calculate) via the Fourier transforms:

\begin{equation}
\int {{{d^3q} \over {(2\pi )^3}}e^{-i\vec q.\vec r}}
\sqrt{1 \over {\vec q \kern 2pt}^2} ={1 \over {2\pi^2 r^2}}
\end{equation} and
\begin{equation}
\int {{{d^3q} \over {(2\pi )^3}}e^{-i\vec q.\vec r}}
\ln \left( { {\vec q \kern 2pt}^2 \over \mu ^2} \right) =-{1
\over {2\pi^2r^3}}
\end{equation}

\noindent where we have slipped back into the simplifying habit of
picking units where $\hbar = c = 1$.

Now we note something very important, namely that the form of
these terms in momentum space are quite special, in that they are
non-analytic. If gravity were somehow renormalizable, then like
other renormalizable theories, all the divergent terms must be
analytic in order to be accommodated back into the starting
Lagrangian. So one has some reason to suspect that the
non-analytic contributions with be finite. In practice, one
does not worry about the ability to renormalize or not, but
simple goes about determining these terms in the hope that
they will be entirely divergent free.

A second piece of magic occurs, for on dimensional grounds alone
we saw the appearance of the renormalization group parameter
$\mu$. This might be expected to undermine the predictive power
of the calculation. But we see the disappearance of this factor
in the move to configuration space. This is very particular to
this calculation and would not be the case for higher order
corrections, i.e. when looking at the first quantum correction to
the relativistic potential. To go any further, one is compelled
discuss renormalization and it's problems in the context of
Einstein gravity.

It might be remarked that much of the strain of doing the tensor
algebra can be overcome by submitting these parts to the
computer; the intermediate results typically reaching into the
thousands of terms, even at one loop.

We should perhaps remind ourselves that it is being intrinsically
assumed that the complete theory of quantum gravity does not
alter this special result. The fact that it does not involve
renormalization gives us reason to believe that the act of full
quantization will not change things, unless a modification to the
classical theory itself is necessary for quantization.

\subsection{The problem with traditional quantization (a
lightning review)}

Those that know the traditional methods and problems with
quantizing are not in need of an introductory review, while those
that have not tinkered with the guts cannot really be properly
shown the approach in an hour. It is with this contradiction in
mind that we set out on a lightning tour of the problem.

The normal approach is to start with a classical theory and try
to quantize it. In reality the world is quantum, and the
classical view is just a special limit. In this sense we are
starting from the top and trying to work down to the more
fundamental. It is an approach fraught with dangers, but it is
the best we can do for now. There is no guarantee that the
attempt to derive the more general from the special case will
bear fruit, and the fact that Einstein gravity can't be quantized
should not be taken to imply that there is no quantum Einstein
gravity.

We have been quantizing non-relativistic systems with success for
some time now, but relativistic theories seem to demand the use
of field theories (to allow for particle creation and
destruction). However, the infinite number of degrees of freedom
tends to be accompanied by infinite quantities in the theory, and
this, very crudely, is the source of the problem when quantizing
field theories. However, some field theories are quantizable,
despite the presence of these infinities. It turns out that in
some very 
special cases it is possible to re-absorb the
infinities into the coupling constants of the original, starting
(classical) theory. It is a mathematical technique that is
difficult to interpret physically; but despite this difficulty it
leads to very good physical predictions for most of the forces of
nature (the electric, weak and strong forces). The fact that this
so called process of renormalization is only successful for a
small class of theories is what makes it predictive. In fact, it
is so successful that it has permitted the unification of the
electric and weak forces and even has a lesser constrained
proposal for also uniting the strong nuclear force. However,
the one remaining force, namely gravitation, does not succumb to
quantization so easily. This suggests the need for something
novel, but should not be taken as reason to totally abandon the
past thinking as a complete failure, and so devoid of
usefulness. Despite this need for change, there seems to be a
proposal for quantizing gravity that is unexpectedly conservative
in its lack novelty. In fact, we will be so traditional as to
investigate the perturbative quantization of gravity. This means
that much of the presentation can be versed in the now rather old
language of field theory, and we can embark on a more detailed
investigation of the problems obstructing the traditional
quantization of Einstein gravity.

\subsubsection{Infra-red divergences}

A common way to deal with the infra-red divergences of a
theory is to give the massless particles a small rest mass,
with the intention of eventually taking the mass goes to zero
limit. This is not an option for Einstein gravity where the
zero mass case and the zero mass limit give differing results
[van Nieuwenhuizen, 1973].

\section{The Perturbative Quantization of Einstein Gravity}

The usual scheme of field quantization is plagued by divergences,
but in some special cases those infinities can be consistently
ploughed back into the theory to yield a finite end result with a
small number of arbitrary constants remaining; these then being
obtained from experiment [Ramond, 1990; Collins, 1984]. This is
the renown scheme of renormalization, disapproved of by some, but
reasonably well defined and yielding results in excellent
agreement with nature. For those disturbed by the appearance of
infinities, there now exists a finite perturbative version
employing analytic continuation (a generalization of the Zeta
function, one loop, technique), that goes under the deceptive
name of `operator regularization'. The fact that after
renormalization some factors, such as mass and charge, are left
undetermined should perhaps not be viewed as a predictive
shortcoming, since the fundamental units of nature are relative.
That is to say, the choice of reference unit (be it mass, length,
time, or charge) is always arbitrary, and then everything else can
be stated in terms of these units. In this sense the final theory
of everything should not, and cannot, predict all.

The fact that only some theories are renormalizable has the
beneficial effect of being selective, and so predictive. This
follows the line of reasoning that physics is more than {\it
descriptive}, but {\it predictive} by virtue of being limited by
the requirement of self consistency.

Unfortunately, in the usual sense, general relativity is {\it
not} renormalizable [Veltman, 1976], and we will run quickly over
the failure of Einstein gravity to quantize perturbatively, by
considering the example of a massive scalar field with gravity.
The starting theory in Euclidean space would be characterized by:

\begin{equation} L = -\sqrt{g}
\left( R + 
\textstyle{1 \over 2} g^{\mu \nu}
\partial_\mu \phi \partial_\nu \phi + 
\textstyle{1 \over 2} m^2 \phi^2
\right)
\end{equation}

\rightline{ \small \it  (using units where $16 \pi G = 1$, $c =
1$)}

One discovers, upon perturbatively quantizing both the matter and
gravitational fields, that the infinities cannot be accounted
for in the traditional way by altering the coefficients of the
terms in the original, starting, Lagrangian. The required counter
terms fall `outside' the original Lagrangian, and so the
infinities cannot be reabsorbed.

One natural thought might be to generalize Einstein gravity by
extending the starting Lagrangian to accommodate the anticipated
counter terms. Here symmetry can be employed as a guide, for by
using the most general starting Lagrangian consistent with the
original symmetries one arranges that the counter terms (which
also retain the symmetry in the absence of a quantum anomaly) fall
back within the Lagrangian. One would not anticipate an anomaly,
as these arise from a quantum conflict between two or more
symmetries, when one must choose between one or the other. Thus
one is lead to the infinitely large starting Lagrangian:

\begin{equation} L_0 = -\sqrt{g_0} 
\left(\matrix { -2\Lambda_0 + R_0 + 
\textstyle{1 \over 2} p_0^2 + 
\textstyle{1 \over 2} m_0^2 \phi_0^2 +
\textstyle{1 \over 4!} \phi_0^4
\lambda_0(\phi_0^2) + p_0^2 \phi_0^2
\kappa_0(\phi_0^2) + R_0 \phi_0^2 \gamma_0(\phi_0^2) \cr
\cr + p_0^4 a_0 (p_0^2,\phi_0^2) + R_0 p_0^2 b_0(p_0^2,\phi_0^2)
+ R_0^2 c_0(p_0^2,\phi_0^2) + R_{0\mu\nu} R_0^{\mu\nu}
d_0(p_0^2,\phi_0^2) + ... }
\right)
\end{equation}

\noindent where $p_0^2$ is shorthand for $g_0^{\mu \nu}
\partial_\mu \phi_0 \partial_\nu \phi_0$ and not the independent
variable of Hamiltonian mechanics.
$\lambda_0$, $\kappa_0$, $\gamma_0$, $a_0$,
$b_0$, $c_0$, $d_0$ ... are arbitrary analytic functions, and the
second line carries all the higher derivative terms. Strictly
this is formal in having neglected gauge fixing and the resulting
presence of ghost particles.

The price for having achieved `formal' renormalization, is that
the theory (with its infinite number of arbitrary renormalized
parameters) is devoid of predictive content. The failure to
quantize has been rephrased from a problem of
non-renormalizability to one of non-predictability.

Despite this, after renormalization we are lead to:

\begin{equation} L = -\sqrt{g} 
\left(\matrix { -2\Lambda + R + 
\textstyle{1 \over 2} p^2 + 
\textstyle{1 \over 2} m^2 \phi^2 +
\textstyle{1 \over 4!} \phi^4 \lambda(\phi^2) + p^2 \phi^2
\kappa(\phi^2) + R \phi^2 \gamma(\phi^2) \cr \cr + p^4 a
(p^2,\phi^2) + R p^2 b(p^2,\phi^2) + R^2 c(p^2,\phi^2) +
R_{\mu\nu} R^{\mu\nu} d(p^2,\phi^2) + ... }
\right)
\end{equation}

This total loss of predictability is a slight exaggeration, as the
people doing the effective field theory of gravity would argue
that the renormalised couplings are of a sane size, so that one
need only a finite number of terms to describe physics at a given
scale. This being the case, there will then only be a finite
number of parameters to determine from experiment, and the
effective theory has a predictive content, albeit not as
strong as one might have hoped for.

But perhaps we can go further than leaving the renormalized
couplings unknown (though `small'), for there remain physical
criterion to pin down some of these arbitrary factors. Since in
general the higher derivative terms lead to acausal classical
behavior, their renormalized coefficient might be put down to
zero on physical grounds. This still leaves the three arbitrary
functions:
$\lambda(\phi^2)$, $\kappa(\phi^2)$ and
$\gamma(\phi^2)$, associated with the terms $\phi^4$,
$p^2 \phi^2$, and
$R \phi^2$ respectively. The last may be abandoned on the grounds
of defying the equivalence principle. To see this, begin by
considering the first term of the Taylor expansion, namely
$R\phi^2$; this has the form of a mass term and so one would be
able to make local measurements of mass to determine the
curvature, and so contradict the equivalence principle (charged
particles, with their non-local fields have this term present
with a fixed coefficient). The same line of reasoning applies to
the remaining terms, $R\phi^4$, $R\phi^6$, ... etc.

This leaves us the two remaining infinite families of ambiguities
with the terms $\phi^4\lambda(\phi^2)$ and
$p^2\phi^2\kappa(\phi^2)$. In the limit of flat space in 3+1
dimensions this will reduce to a renormalized theory in the
traditional sense if $\lambda(\phi^2)=constant$, and
$\kappa(\phi^2)=0$. So one is lead to proposing that the physical
parameters should be:

\begin{equation}
\matrix {
\Lambda = \kappa(\phi^2) = \gamma(\phi^2) = 0 \cr \cr
a(p^2,\phi^2) = b(p^2,\phi^2) = c(p^2,\phi^2) = d(p^2,\phi^2)
=... =0 \cr \cr
\lambda(\phi^2) = \lambda = {\it scalar\ particle\ self\
coupling\ constant} \cr \cr m = {\it mass\ of\ the\ scalar\
particle} }
\end{equation}

\noindent and so the renormalized theory of quantum gravity for a
scalar field should have the form:

\begin{equation} L = -\sqrt{g} 
\left( -2\Lambda + R + 
\textstyle{1 \over 2} g^{\mu \nu}
\partial_\mu \phi \partial_\nu \phi + 
\textstyle{1 \over 2} m^2 \phi^2 +
\textstyle{1 \over 4!} \lambda \phi^4
\right)
\end{equation}

One might now worry about the renormalization group pulling the
coupling constants around. This is an open point to which I feel
one of several things might happen:

\begin{description} { \it

\item \hskip .4cm $\bullet$ The couplings, set to
zero at a low energy scale, might reappear around the Plank
scale. Whether the resulting theory then makes sense is a matter
for dispute.

\item \hskip .4cm $\bullet$ Certain coupling
constants (beyond those already set to zero) should be related,
in order that the beta functions of the zeroed couplings be zero
(a fixed point), so ensuring that all their couplings remain at
zero. This consistency condition could be the basis of a
unification scheme, although its implementation might not be
possible within the perturbative formulation. }
\end{description}

This is a highly non-trivial matter that needs looking at more
closely.

\subsection{Regularization method}

On a diverse, but related track, one might wonder which
renormalization scheme to choose for implementing the scheme
proposed above. In this context analytic continuation [Bollini et
al., 1964; Speer, 1968; Salam and Strathdee, 1975; Dowker and
Critchley, 1976; Hawking, 1977] is very appealing in being
finite, and in this context there is an `unsung hero' in the
guise of operator regularization, which I think deserves a
mention [McKeon and Sherry, 1987; McKeon et al., 1987; McKeon et
al., 1988; Mann, 1988; Mann et al., 1989; Culumovic et al., 1990;
Shiekh, 1990].

In operator regularization one avoids the divergences by using
the analytic continuation:

\begin{equation}
\Omega^{-m} = 
\lim \limits_{\varepsilon \to 0} {d^n \over d\varepsilon^n}
\left( {\varepsilon^n \over n! }
\Omega^{-\varepsilon - m}
\right)
\end{equation}

\noindent where $n$ is chosen large enough to eliminate the
infinities (the loop order is sufficient). Actually, operator
regularization is a bit of a misnomer, since it need not be
applied to an operator and does not just regulate, but also
renormalizes all in one. However, under this form of the method
{\it all} theories are finite and predictive (gravity included).

This form of the method locates and eliminates the poles
(automated minimal subtraction). This is clearly seen by how it
treats some terms of the Maclaurin expansion. The taming of
$1$ yields:

\begin{equation}
\mathop {\lim}\limits_{\varepsilon \to 0} {d \over
{d\varepsilon }}\left( {\varepsilon .1} \right)
=1
\end{equation}

\noindent while, on the other hand, the taming of
${1 / \varepsilon}$ yields:

\begin{equation}
\mathop {\lim}\limits_{\varepsilon \to 0} {d \over
{d\varepsilon }}\left( {\varepsilon .
{ 1  \over \varepsilon }
} \right)=0
\end{equation}

This realized, the general form is easily located, and is given
by:

\begin{equation}
\Omega^{-m} = 
\lim \limits_{\varepsilon \to 0} {d^n \over d\varepsilon^n}
\left( (1+\alpha_1 \varepsilon +...+\alpha_n \varepsilon^n)
{\varepsilon^n \over n! }
\Omega^{-\varepsilon - m}
\right)
\end{equation}
\rightline{\small \it (the alphas being ambiguous)}

This form is no longer too powerful, and gravity must again be
dealt with as before, setting most of the final renormalized
parameters to zero on physical grounds

The method of operator regularization has the strength of
explicitly maintaining invariances, further even than dimensional
regularization, for dimension dependent invariances are not
disturbed. It is further blessed with the feature of being finite
throughout, as the Zeta function technique [Salam and Strathdee,
1975; Dowker and Critchley, 1976; Hawking, 1975]. But unlike the
Zeta function method, it is not limited in applicability to one
loop, being valid to all orders.

\subsection{The first few Feynman rules}

To go about a perturbative calculation we should have
the Feynman rules in hand, of which gravity is exceptional
in having an infinite number due to the presence of the square
root in the Lagrangian which makes it non-polynomial. This does
not cause a great problem, because to any finite loop order, only
a finite number of Feynman rules are needed. The first few
rules we list as:
\vskip .3cm

$\bullet$ The graviton propagator:
\vglue .5cm
\hskip .8cm
\epsfbox{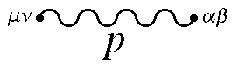}
\vglue -1.5cm

\begin{equation}
{{\eta _{\mu \alpha }\eta _{\nu \beta }+\eta _{\mu
\beta }\eta _{\nu \alpha }-\eta _{\mu \nu }\eta _{\alpha \beta }}
\over {p^2}}
\end{equation}

\vglue .5cm
$\bullet$ The scalar propagator:
\vglue .5cm
\hskip 1cm
\epsfbox{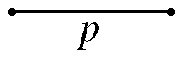}
\vglue -1.5cm

\begin{equation}
{1 \over {p^2+m^2}}
\end{equation}

\vglue .5cm
$\bullet$ The first interaction vertex:
\vglue .5cm
\hskip 1cm
\epsfbox{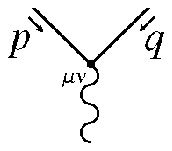}
\vglue -1.5cm

\begin{equation}
{\textstyle{1 \over 2}}\left( {\eta _{\mu \nu }
(p.q-m^2)-p_\mu
q_\nu -p_\nu q_\mu } \right)
\end{equation}

\section{The method in Action}

To see some of the assembled machinery in action,
we can go about calculating a simple one loop
diagram, namely one part of the correction to
the scalar propogator:

\vglue 1cm
\hskip .25cm
\epsfbox{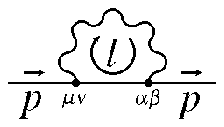}
\vglue -2cm

\begin{eqnarray}
=\displaystyle\int_{-\infty }^\infty {{{d^4l} \over
{(2\pi )^4}}}
&\left( {\displaystyle{ 1 \over {l^2+m^2}}}
\right)\left( 
\displaystyle{{{\eta _{\mu
\alpha }\eta _{\nu \beta }+\eta _{\mu \beta }\eta _{\nu \alpha
}-\eta _{\mu \nu }\eta _{\alpha \beta }} 
\over{(l+p)^2}}}
\right) \nonumber \\
&\times {\textstyle{ 1 \over 2}}
\left( {\eta _{\mu \nu }(p.l-m^2)-p_\mu l_\nu -p_\nu l_\mu }
\right) \\
&\times {\textstyle{1 \over 2}}\left( {\eta _{\alpha \beta
}(p.l-m^2)-p_\alpha l_\beta -p_\beta l_\alpha } \right) \nonumber
\end{eqnarray}

\noindent
Expand out the indices to yield:

\begin{equation}
=\int_{-\infty }^\infty  {{{d^4l} \over {\left( {2\pi }
\right)^4}}}{1 \over {l^2+m^2}}{1 \over {\left( {l+p}
\right)^2}}\left( {p^2l^2+2m^2p\cdot l-2m^4} \right)
\end{equation}

\noindent
and then introduce the standard Feynman parameter `trick':

\begin{equation}
{1 \over
{D_1^{a_1}D_2^{a_2}...D_k^{a_k}}}={{\Gamma (a_1+a_2+...a_k)}
\over {\Gamma (a_1)\Gamma (a_2)...\Gamma (a_k)}}\int_0^1
{...}\int_0^1 {dx_1...dx_k}{{\delta
(1-x_1-...x_k)x_1^{a_1-1}...x_k^{a_k-1}} \over {\left(
{D_1x_1+...D_kx_k} \right)^{a_1+...a_k}}}
\end{equation}

\noindent
to yield:

\begin{equation}
=\int_{-\infty }^\infty  {{{d^4l} \over {\left(
{2\pi }
\right)^4}}\int_0^1 {dx}}{{p^2l^2+2m^2p\cdot l-2m^4} \over {\left[
{l^2+m^2x+p^2\left( {1-x} \right)+2l\cdot p\left( {1-x} \right)}
\right]^2}}
\end{equation}

Remove divergences using operator regularization:

\begin{equation}
\Omega ^{-m}=\mathop {\lim}\limits_{\varepsilon
\to 0}{{d^n}
\over {d\varepsilon ^n}}\left( {\left( {1+\alpha _1\varepsilon
+...+\alpha _n\varepsilon ^n} \right){{\varepsilon ^n} \over
{n!}}\Omega ^{-\varepsilon -m}} \right)
\end{equation}

\noindent
$n$ being chosen sufficiently large to cancel the infinities. For
the case in hand $n=1$ is adequate.

\begin{equation}
\Omega ^{-2}=\mathop {\lim}\limits_{\varepsilon
\to 0}{d \over {d\varepsilon }}\left( {\left( {1+\alpha
\varepsilon }
\right)\varepsilon \Omega ^{-\varepsilon -2}} \right)
\end{equation}

This leads to:

\begin{equation}
=\int_0^1 {dx}\;\mathop
{\lim}\limits_{\varepsilon \to 0}{d
\over {d\varepsilon }}\int_{-\infty }^\infty  {{{d^4l} \over
{\left( {2\pi } \right)^4}}}\left( {\varepsilon \left( {1+\alpha
\varepsilon } \right){{p^2l^2+2m^2p\cdot l-2m^4} \over {\left[
{l^2+m^2x+p^2\left( {1-x} \right)+2l\cdot p\left( {1-x} \right)}
\right]^{\varepsilon +2}}}} \right)
\end{equation}

Then performing the momentum integrations using [Ramond, 1990]:

\begin{equation}
\int_{-\infty }^\infty  {{{d^{2\omega }l} \over
{\left( {2\pi }
\right)^{2\omega }}}}{1 \over {\left( {l^2+M^2+2l\cdot p}
\right)^A}}={1 \over {(4\pi )^\omega \Gamma (A)}}{{\Gamma (A-\omega
)} \over {(M^2-p^2)^{A-\omega }}}
\end{equation}

\begin{equation}
\int_{-\infty }^\infty  {{{d^{2\omega }l} \over
{\left( {2\pi }
\right)^{2\omega }}}}{{l_\mu } \over {\left( {l^2+M^2+2l\cdot p}
\right)^A}}=-{1 \over {(4\pi )^\omega \Gamma (A)}}p_\mu {{\Gamma
(A-\omega )} \over {(M^2-p^2)^{A-\omega }}}
\end{equation}

\begin{equation}
\int_{-\infty }^\infty  {{{d^{2\omega }l} \over
{\left( {2\pi }
\right)^{2\omega }}}}{{l_\mu l_\nu } \over {\left( {l^2+M^2+2l\cdot
p} \right)^A}}={1 \over {(4\pi )^\omega \Gamma (A)}}\left[ {p_\mu
p_\nu {{\Gamma (A-\omega )} \over {(M^2-p^2)^{A-\omega }}}+{{\delta
_{\mu \nu }} \over 2}{{\Gamma (A-\omega -1)} \over
{(M^2-p^2)^{A-\omega -1}}}} \right]
\end{equation}

\noindent
yields the finite expression:

\begin{equation}
={1 \over {\left( {4\pi } \right)^2}}\int_0^1
{dx}\;\mathop
 {\lim}\limits_{\varepsilon \to 0}{d \over {d\varepsilon }}
\left(
{{{\varepsilon \left( {1+\alpha \varepsilon } \right)} \over
{\Gamma (\varepsilon +2)}}\left( 
\matrix{{{\textstyle p^4(1-x)^2\Gamma
(\varepsilon )} \over 
{\left[ {\textstyle m^2x+p^2x\left( {1-x} \right)}
\right]^\varepsilon }}
+2{{\textstyle p^2\Gamma (\varepsilon -1)} \over {\left[
{\textstyle m^2x+p^2x\left( {1-x} \right)} \right]^{\varepsilon -1}}}\cr
  -2{{\textstyle m^2p^2(1-x)\Gamma (\varepsilon )} \over {\left[
{\textstyle m^2x+p^2x\left( {1-x} \right)} \right]^\varepsilon }}
-2{{\textstyle m^4\Gamma
(\varepsilon )} \over {\left[ {\textstyle m^2x+p^2x\left( {1-x} \right)}
\right]^\varepsilon }}\cr} \right)} \right)
\end{equation}

Doing the $\varepsilon $ differential using:

\begin{equation}
\mathop {\lim}\limits_{\varepsilon \to 0}{d \over
{d\varepsilon }}\left( {{{\varepsilon \left( {1+\alpha
\varepsilon } \right)}
\over {\Gamma (\varepsilon +2)}}\left( {a{{\Gamma (\varepsilon )}
\over {\chi ^\varepsilon }}+b{{\Gamma (\varepsilon -1)} \over {\chi
^{\varepsilon -1}}}} \right)} \right)=-a+\left( {a-b\chi }
\right)\left( {\alpha -\ln (\chi )} \right)
\end{equation}

\noindent
yields:

\begin{equation}
={1 \over {\left( {4\pi } \right)^2}}\int_0^1
{dx}\left(
\matrix{\left( {\left( {2m^4+2m^2p^2-p^4} \right)+p^4x\left( {4-3x}
\right)} \right)\left( {\ln \left( {m^2x+p^2x(1-x)} \right)-\alpha
} \right)\cr \cr
  +2m^4+2m^2p^2-p^4-p^2x\left( {2m^2-2p^2+p^2x} \right)\cr}
\right)
\end{equation}

\noindent
and finally performing the $x$ integration gives rise to the final
result in Euclidean space, namely:

\vglue 1cm
\hskip .5cm
\epsfbox{oneloop.eps}
\vglue -1.7cm

\begin{equation}
={{m^4} \over {\left( {4\pi} \right)^2}}
\left( \matrix{\left(
{3+2{{p^2} \over {m^2}}+{{m^2} \over {p^2}}} \right)\ln (1+{{p^2}
\mathord{\left/ {\vphantom {{p^2} {m^2}}} \right.
\kern-\nulldelimiterspace} {m^2}})-1-{5 \over 2}{{p^2} \over
{m^2}}
\cr
-{1 \over 6}{{p^4} \over {m^4}}+2\left( {1+{{p^2} \over {m^2}}}
\right)\left( {\ln ({{m^2} \mathord{\left/ {\vphantom {{m^2} 
{\mu
^2}}} \right. \kern-\nulldelimiterspace} {\mu ^2}})-\alpha }
\right)\cr} \right)
\end{equation}

\noindent where there is no actual divergence at $p=0$, and it
should be commented that the use of a computer mathematics
package can in general greatly reduced `calculator' fatigue.
The renormalization group factor $\mu $ appears on dimensional
grounds.

In general the Feynman rules are large and the tensor
algebra immense. Much of the simplicity is restored by
submitting this part of the complexity to the computer. Even
so, the intermediate results can be so extensive that even a
super-computer can choke without trivial help. For example,
imagine one had the contraction of three tensors:

\begin{equation}
\alpha _{\mu \nu }\beta _{\rho \sigma }
\gamma ^{\mu \nu \rho \sigma }
\end{equation}

\noindent each of which consists of many terms. Then the
computer, in trying to contract out the indices, tends to
expand out the entire expression which can easily lead to
thousands of terms, that can overpower the computers memory.

The resolution lies in the trivial step of asking
the computer to initially expand out only $\alpha$
for example:

\begin{equation}
({\alpha _1}_{\mu \nu }+{\alpha _2}_{\mu \nu }+\ldots )
\beta_{\rho \sigma }\gamma ^{\mu \nu \rho \sigma }
\end{equation}

In this way the computer is presented with several
terms that can each be contracted separately. This
seemingly innocuous move can make all the difference
between the computer being able to perform the calculation
and not. It is details like this that in practice can
occupy much of the investigators time.

\subsection{Discussion}

We are now left with a finite theory that has few arbitrary
constants, and so is predictive. Despite the present lack of
experimental data to test it against, and regardless of the patch
work line of reasoning invoked to arrive at this hypothesis, one
might alter perspective and simply be interested in investigating
the consequences of such a scheme for its own sake, where many of
the arbitrary factors have been set to zero, for whatever reason.
At this stage any well behaved, finite theory, is worth
investigating; and it is unfortunate that we don't have the
guiding hand of mother nature to assist us in this guessing
game.\\

{\Large \bf Acknowledgments}

My gratitude to John Strathdee and Seifallah Randjbar
Daemi for listening to and commenting upon, if not necessarily
agreeing with, my ideas. Special thanks are extended to Gerry
McKeon and Roberto Percacci for a responsive ear and
constructive criticism.

As always, many thanks to the organizers of the conference for
having once again invited me to participate.

Special thanks to Migdal Virasoro, who made sure I was able
to go and present my review. \\

{\Large \bf References}

\begin{description} {\small

\item \hskip .4cm $\bullet$ {\bf C.J. Isham}, {\it `Quantum
Gravity - An Overview'}, in ``Quantum Gravity 2: A Second Oxford
Symposium'', pp. 1-62, eds. C.J. Isham, R. Penrose and D.W.
Sciama, (Oxford University Press, Oxford, 1981).

\item \hskip .4cm $\bullet$ {\bf A.Y. Shiekh},  {\it `The
Perturbative Quantization of Gravity'}, in  ``Problems on High
Energy Physics and Field Theory'', Proceedings of the XVII
workshop 1994, pp. 156-165, Protvino, 1995.
\\ {\bf A.Y. Shiekh},  {\it `Quantizing Orthodox Gravity'}, Can.
J. Phys.,  {\bf 74}, 1996, 172-.

\item \hskip .4cm $\bullet$ {\bf J.F. Donoghue},  
{\it `Leading Quantum Corrections to the Newtonian Potential'},
Phys. Rev. Lett.,  {\bf 72}, 1994, 2996-.
\\ {\bf J.F. Donoghue},  
{\it `General relativity as an effective field theory:
The leading quantum corrections'}, Phys.
Rev.,  {\bf D50}, 1994, 3874-.

\item \hskip .4cm $\bullet$ {\bf R.P. Feynman}, {\it ``Lectures
on Gravitation''} (Addison-Wesley, 1995).
\\ {\bf T.W.B. Kibble}, {\it `Is a Semi-Classical Theory of
Gravity Viable?'}, in ``Quantum Gravity 2: A Second Oxford
Symposium'', pp. 63-80, eds. C.J. Isham, R. Penrose and D.W.
Sciama, (Oxford University Press, Oxford, 1981).

\item \hskip .4cm $\bullet$ {\bf P. Ramond}, {\it ``Field
Theory: A Modern Primer''},
 2nd Ed (Addison-Wesley, 1990).
\\ {\bf J. Collins}, {\it ``Renormalization''}, (Cambridge
University Press, London, 1984).

\item \hskip .4cm $\bullet$ {\bf P. van Nieuwenhuizen},
{\it `Radiation of Massive Gravitation'}, Phys. Rev.,
{\bf D7}, 1973, 2300-.

\item \hskip .4cm $\bullet$ {\bf M.J.G. Veltman}, {\it `Quantum
Theory of Gravitation'}, Les Houches XXVIII, ``Methods In Field
Theory'', pp. 265-327, eds. R. Ballan and J. Zinn-Justin,
(North-Holland, Amsterdam, 1976).

\item \hskip .4cm $\bullet$ {\bf C. Bollini, J. Giambiagi and
A. Dom\`{\i}nguez}, {\it `Analytic Regularization and the
Divergences of Quantum Field Theories'}, Nuovo Cimento {\bf 31},
1964, 550-.
\\ {\bf E. Speer}, {\it `Analytic Renormalization'}, J. Math.
Phys. {\bf 9}, 1968, 1404-.
\\ {\bf A. Salam and J. Strathdee}, {\it `Transition
Electromagnetic Fields in Particle Physics'}, Nucl. Phys. {\bf
B90}, 1975, 203-.
\\ {\bf J. Dowker and R. Critchley}, {\it `Effective Lagrangian
and energy-momentum tensor in de sitter space'}, Phys. Rev., {\bf
D13}, 1976, 3224-. 
\\ {\bf S. Hawking}, {\it `Zeta Function Regularization of Path
Integrals in Curved Space'}, Commun. Math. Phys., {\bf 55}, 1977,
133-.

\item \hskip .4cm $\bullet$ {\bf D. McKeon and T. Sherry}, {\it
`Operator Regularization of Green's Functions'}, Phys. Rev.
Lett., {\bf 59}, 1987, 532-. 
\\ {\bf D. McKeon and T. Sherry}, {\it `Operator Regularization
and one-loop Green's functions'}, Phys. Rev., {\bf D35}, 1987,
3854-.
\\ {\bf D. McKeon, S. Rajpoot and T. Sherry}, {\it `Operator
Regularization with Superfields'}, Phys. Rev., {\bf D35}, 1987,
3873-.
\\ {\bf D. McKeon, S. Samant and T. Sherry}, {\it `Operator
regularization beyond lowest order'}, Can. J. Phys., {\bf 66},
1988, 268-. 
\\ {\bf R. Mann.}, {\it `Zeta function regularization of Quantum
Gravity'}, In Proceedings of the cap-nserc Summer Workshop on
Field Theory and Critical Phenomena. Edited by G. Kunstatter, H.
Lee, F. Khanna and H. Limezawa, World Scientific Pub. Co. Ltd.,
Singapore, 1988, p. 17-. 
\\ {\bf R. Mann, D. McKeon, T. Steele and T. Tarasov}, {\it
`Operator Regularization and Quantum Gravity'}, Nucl. Phys., {\bf
B311}, 1989, 630-. 
\\ {\bf L. Culumovic, M. Leblanc, R. Mann, D. McKeon and T.
Sherry}, {\it `Operator regularization and multiloop Green's
functions'}, Phys. Rev., {\bf D41}, 1990, 514-.
\\ {\bf A. Shiekh}, {\it `Zeta-function regularization of quantum
field theory'}, Can. J. Phys., {\bf 68}, 1990, 620-.

}
\end{description}

\end{document}